\documentclass[twocolumn,showpacs,preprintnumbers,amsmath,amssymb]{revtex4}         \vskip5mm

\begin{document}
\title{ CKM matrix unitarity and a novel type of global fits}
\author{Petre Di\c t\u a}
\affiliation{ Institute of Physics and Nuclear Engineering,
P.O. Box MG6, Bucharest, Romania}

\begin{abstract}
The aim of the paper is to propose one paradigm change of CKM global fits on experimental data from electroweak sector. The change refers to using in fits the exact unitarity constraints expressed in terms of four invariant parameters, such as  moduli of the CKM  matrix, and to take into account an important set  of the available  experimental data. In the paper we use
 data  from nuclear beta decays, and  from leptonic and semileptonic decays, in order to  find the most probable numerical form of the CKM matrix, as well as the determination of decay constants, $f_{P}$, and of various form factors  $f_+^{Pp}(0)$, directly from experimental results.
\end{abstract}


\maketitle

 The  consistency problem of  experimental data with unitarity constraints  was recently solved, and a procedure for recovering the  CKM matrix elements from error affected data was provided in \cite{PD}. These unitarity constraints say that the four independent parameters  $s_{ij}$ and $\cos\delta$, with the standard notation,  should take physical values, i.e. $s_{ij}\in (0,1)$ and $\cos\delta \in (-1,1)$, when they are obtained from the equations: 
\begin{eqnarray}
V_{ud}^2&=&c^2_{12} c^2_{13},\,\, V_{us}^2=s^2_{12}c^2_{13},\,\,V_{ub}^2=s^2_{13}\nonumber \\
 V_{cb}^2&=&s^2_{23} c^2_{13},\,\,
 V_{tb}^2=c^2_{13} c^2_{23},\nonumber\\
V_{cd}^2&=&s^2_{12} c^2_{23}+s^2_{13} s^2_{23} c^2_{12}+2 s_{12}s_{13}s_{23}c_{12}c_{23}\cos\delta,\nonumber\\
V_{cs}^2&=&c^2_{12} c^2_{23}+s^2_{12} s^2_{13} s^2_{23}-2 s_{12}s_{13}s_{23}c_{12}c_{23}\cos\delta,~~~~~\label{uni}\\
V_{td}^2&=&s^2_{13}c^2_{12}c^2_{23}+s^2_{12}s^2_{23}-2 s_{12}s_{13}s_{23}c_{12}c_{23}\cos\delta\nonumber,\\
V_{ts}^2&=&s^2_{12} s^2_{13} c^2_{23}+c^2_{12}s^2_{23} +2 s_{12}s_{13}s_{23}c_{12}c_{23}\cos\delta\nonumber
\end{eqnarray}
where $V_{ij}=|U_{ij}|$, and $U_{ij}$ are the entries of the CKM matrix. It was shown in \cite{PD} that if the physical quantities could also depend upon  CKM matrix moduli, the reconstruction of a unitary matrix from such data is essentially unique,  and in the following the used independent parameters will be the moduli $V_{ij}$. For example, if $V_{us}=a,\,V_{ub}=b,\,V_{cd}=d,\,{\rm and}\,\,V_{cb}=c$ is one  set of four independent moduli, the relations (\ref{uni}) give the following solution
\begin{eqnarray}
s_{13}=V_{ub}=b,\,\,s_{12}=\frac{a}{\sqrt{1-b^2}}, \,\,s_{23}=\frac{c}
{\sqrt{1-b^2}}\label{sol}\end{eqnarray}
\begin{eqnarray}
&&\cos\delta= \label{col}\\
&&\frac{(1-b^2)(d^2(1-b^2)-a^2)+c^2(a^2+b^2(a^2+b^2-1))}{2 a b c \sqrt{1-a^2-b^2}\sqrt{1-b^2-c^2}}\nonumber
\end{eqnarray}
Because there are 58 groups of four independent moduli, one gets 165 different expressions for $\cos\delta$ and all have to take (roughly) the same value when computed from Eqs. (\ref{uni}). Thus the $\chi^2$-function 
have to contain two kinds of terms: the first has to impose the fulfillment of unitarity constraints, and the second should take into account the physical quantities measured in experiments. 

Concerning the experimental data we will consider  data coming from super-allowed $0^+ \rightarrow 0^+$ nuclear beta decays, and from  leptonic and semileptonic decays.

In the standard model the decay rate for purely leptonic decay is given by 
\begin{eqnarray}
\Gamma\,(P\rightarrow \ell\,\overline{\nu}_{\ell})=\frac{G_F^2}{8\,\pi}|U_{q q'}|^2\,f_P^2\,M_P\,m_{\ell}^2\left(1-\frac{m_{\ell}^2}{M_P^2}\right)^2\label{g1}
\end{eqnarray}
where $G_F$ is the Fermi constant, $M_P$ and $m_{\ell}$ are the masses of the decaying meson, and, respectively, of the final lepton, $U_{qq'}$ is the corresponding CKM matrix element,  and $f_P $ is the decay constant. In general  one has  also to take into account the radiative corrections which lead to a minor modification of the above formula.

 The physical observable for  semileptonic decays, that depends on $|U_{qq'}|$ and  $f(q^2)$,  is the differential decay rate which under assumption  of massless leptons is  written as
\begin{eqnarray}
\frac{d\,\Gamma(H\rightarrow P\,\ell\,\nu_{\ell})}{d q^2}=\frac{G_F^2\,|U_{qq'}|^2}{192\pi^3 M_H^3}\lambda^{3/2}(q^2)|f(q^2)|^2\label{g3}
\end{eqnarray}
where $q=p_H-p_P$ is the transferred momentum, and
\begin{eqnarray}\lambda(q^2)=(M_H^2+M_P^2-q^2)^2-4 M_H^2 M_P^2 \end{eqnarray} is the usual triangle function,  and $f(q^2)$ is  the global form factor which is a combination of $f_+(q^2)$ and $f_-(q^2)$. The experimenters provide numerical values for  products of the form $|U_{qq'}\,f_+(0)|$, and  in this paper we will use these numerical values.
Hence from such  experiments one measures, up to known factors,   products of the form
\begin{eqnarray}|U_{ij}|^2\times f_P^2,\,\, {\rm and/or},\,\,\,
 |U_{ij}|^2\times |f(q^2)|^2\label{p1}\end{eqnarray}
 It is clear that from  such measurements one cannot find {\em two  unknown} quantities, let's say, $|f(q^2)|$ and $|U_{ij}|$, if we have no supplementary constraints. Our point of view is that the unitarity constraints, which depend only on $ |U_{ij}| $ moduli, see relations (\ref{sol})-(\ref{col}), provide the necessary  tool for the separation of  moduli, and   $f(q^2)$, or $f_P$.

 Before  defining our  type of global fit we make one natural assumption, which is:
\textit {the numerical values for all the measured moduli,  $|U_{ij}|,$ must be the same irrespective of the physical processes used to determine them.} The other parameters, such as the decaying constants  $f_P$, form factors $f_+(0)$, $g_A/g_V$, etc., which parametrize the data from each given experiment, are considered free parameters to be found from  fit.
 
The     first   piece containing  unitarity constraints entering the $\chi^2$-function  has the  form
\begin{eqnarray}
&&\chi^2_{1}=\nonumber\\
&&\sum_{j=u,c,t}\left(
\sum_{i=d,s,b}V_{ji}^2-1\right)^2
+\sum_{j=d,s,b}\left(
\sum_{i=u,c,t}V_{ij}^2-1\right)^2~~\label{chi1}\nonumber\\
&& + \sum_{i < j}(\cos\delta^{(i)} -\cos\delta^{(j)})^2,\,\,\,\,-1\le\cos\delta^{(i)}\le 1
\end{eqnarray}
and the second component, 
which takes into account the experimental data,  is
\begin{eqnarray}
\chi^2_2=\sum_{i}\left(\frac{d_{i}-\widetilde{d}_{i}}{\sigma_{i}}\right)^2\label{chi2} 
\end{eqnarray}
where  $d_{i}$ are the theoretical functions one wants to be found from fit, $\widetilde{d}_{i}$ is  the numerical matrix that describes the corresponding experimental data, while  $\sigma$ is the matrix of errors associated to $\widetilde{d}_{i}$. 
 In the following our  $\chi^2$-function will be 
\begin{eqnarray}
\chi^2=\chi^2_1 +\chi^2_2\end{eqnarray}

Concerning experimental data we use the following.  Knowledge on  $|U_{ud}|$ comes mainly from three different sources: a) super allowed, $0^+\rightarrow 0^+$, nuclear beta decays, see \cite{HT}, \cite{GS}, and  \cite{HT1}, b) neutron beta decay, $n\rightarrow p\, e^+ \nu$, see \cite{AS}-\cite{PS}, c) and    pion beta decay $\pi^+\rightarrow \pi^0 e^+\nu$, \cite{DP}.

The used data  for the determination of the decay constants $f_{\pi},\,\,f_K,\,\,f_B,\,\,f_D,$  $\,\, {\rm and}\,\,f_{D_s^+}$ are from the papers  \cite{WM}, \cite{IK}, \cite{Ba}, \cite{AR}, and, respectively, from \cite{BA}-\cite{ME}. Numerical results on $|f_+^{K\pi}(0)U_{us}|$ are from the papers \cite{JR}-\cite{FA1}, those upon $|f_+^{B\pi}(0)U_{ub}|$ come from \cite{JM} and \cite{Bau}, and the ratio $|U_{cd}\,f_+^{D\pi}(0)/f_+^{DK}(0)\, U_{cs}|$ is given in  \cite{JML} and  \cite{GSH}. The papers \cite{DB}-\cite{AB2} provide data on $|{\cal F}(1)U_{cb}|$, and  \cite{DB1} and \cite{MAt}-\cite{KAb1} provide values for $|G(1)U_{cb}|$.

The central values and uncertainties used in  fit are those published in the above papers, and we combined the statistical and systematic uncertainties in quadrature when experimenters provided both of them.

According to \cite{HT}, the super-allowed beta decays between $T=1$ analog $0^+$ states, together with the conserved vector current (CVC) hypothesis, lead to the conclusion  that the $ft$ values should be the same irrespective of the nucleus, i.e.
\begin{eqnarray}
ft=\frac{K}{|G_V|^2\,|M_F|^2}= \textrm{const},\label{g4}
\end{eqnarray}
where $K$ is the vector coupling constant for semi-leptonic weak intercations, $f$ is the statistical rate function, and $t$ is the partial half-life. Because the above relation  is only approximately satisfied, one defines a ``corrected'' $\mathcal{F}t$ value, which should be ``constant'',  as
\begin{eqnarray}
\mathcal{F}t\equiv ft(1-\delta_R)(1-\delta_C)=\frac{K}{2|G_V|^2(1+\Delta_R^V)}\end{eqnarray}
where $\delta_C$ is the isospin-symmetry-breaking correction, $\delta_R $ is the transition-dependent part of the radiative correction, and $\Delta_R^V $ is  the transition-independent part. Numerical values for  ${\cal{F}}t$ are given in  \cite{HT}, \cite{GS}, and \cite{HT1}. In our fit we  use  the above formula with $|G_V|^2= |U_{ud}|^2$, by supposing that $g_V(0)=1$, as CVC requires, and $|U_{ud}| $ and  $\Delta_R^V $ are the free parameters to be obtained from  fit.
 Similarly for the neutron beta decay data we make use of the formula
\begin{eqnarray}
|U_{ud}|^2(1+3\lambda^2)=\frac{4908.7(1.9)\,s}{\tau_n}\\\nonumber
\end{eqnarray}
see \cite{It}, where $\tau_n$ is the neutron mean life, and the free parameters are $|U_{ud}| $ and $\lambda = g_A/g_V$.

If one or more of the above parameters could be measured in other experiments, this   approach allows us  to take the results of these measurements  into account. That is the case of the ratio $g_A/g_V$ which enters in  the  measured asymmetry parameter $A_0$, see  papers \cite{PB}-\cite{PL}. Their  effect was a lowering of $\lambda$ to the value given in the Table, while by using only results from neutron beta decay data the value,  $\lambda=-1.27092\pm 0.00394$, is obtained.

Values and corresponding  uncertainties obtained from the fit are given in  Table.

 The surprising result of our fit was that  $\Delta_R^V$  is not transition-independent as it is usually assumed, see Refs.\,\cite{HT}-\cite{HT1}, and  \cite{WJM}-\cite{AS1}. For example, if one uses the  data on nuclear beta decays from  \cite{HT1},  the  $\Delta_R^V$ variation is from $2.193\%$ for $^{22}$Mg, to $2.579\%$ \,for $^{54}$Co nucleus. For this case  the corresponding mean  value and uncertainty are given in Table.
 
If one makes use of  Savard \textit{et al.} data, \cite{GS},  one gets 
\begin{eqnarray}
 \Delta_R^V=(2.294 \pm 0.131)\%\label{rad1}\end{eqnarray}
 and the values spreading is between $2.027\%$ for $^{74}$Rb, and $2.429\%$ for $^{34}$Cl. 

Our approach allows   the use of all  the seventeen values from  \cite{HT}, and one gets  $\Delta_R^V=(2.362 \pm 2.133)\%$, where  the value of  $\Delta_R^V$ for $^{42}$Ti provided by the fit is negative $\Delta_R^V=-4.673\%$! However the result makes sense since the corresponding ${\cal{F}}t$ is $3300 \pm 1100$ which is far away from the mean value given in \cite{HT} which is around $3072$. Hence the fit  suggests us to throw out this value. By excluding also the $^{18}$Ne and  $^{30}$S data, that lead to greater values than the mean by a factor of 3, and, respectively of 2,  one obtains 
\begin{eqnarray} \Delta_R^V=(2.364 \pm 0.182)\%\label{rad2}\end{eqnarray}
Looking at the three $ \Delta_R^V$ values,  that from the Table, and  those obtained by using data from \cite{GS}, Eq. (\ref{rad1}), and, respectively, from 
\cite{HT}, Eq. (\ref{rad2}), one observes that they are compatible within the errors, and the better data are those coming from Ref.\,\cite{GS}. In all these three cases the errors provided by the fit are bigger than the theoretical estimates. We remind that the theoretical estimates are
\begin{eqnarray} \Delta_R^V(old)&=&(2.40 \pm 0.08)\%,\quad {\rm and}\\
\Delta_R^V(new)&=&(2.361 \pm 0.038)\%\label{rad3}\end{eqnarray}
given respectively, in Ref.\,\cite{AS1}, and \cite{HT1}. 

\[ \begin{array}{||c|l||}\hline\hline
&\\
\mbox{~Parameters~~} & ~ \mbox{Central Values and Errors ~~}\\
&\\\hline
&\\
V_{ud} &~~ 0.974022 \,\pm \,3.9 \times 10^{-6}\\
V_{us} &~~ 0.226424 \,\pm \, 3.9\times 10^{-6}\\
V_{ub} &~~ (3.57604   \,\pm \, 0.00002)\times10^{-3}~~~\\
V_{cd} &~~ 0.226261  \,\pm  \,3.9\times 10^{-6}\\
V_{cs} &~~ 0.973324 \,\pm\, 4.1\times10^{-6}\\
V_{cb} &~~ (38.0239  \pm 0.0002)\times10^{-3} \\
V_{td} &~~ (9.28657  \pm 0.000035)\times10^{-3} \\
V_{ts} &~~ (37.0454  \pm 0.0002)\times10^{-3}\\
V_{tb} &~~ 0.999270  \pm  2.2\times10^{-7}\\
\Delta_R^V & ~~(2.399\pm 0.108)\%\\
g_A/g_V & ~~ -1.26924  \pm 0.00510\\
\delta_c & ~~ (3.104\pm 0.096)\%\\
f_{\pi} & ~~ 130.784  \pm 1.323\\
f_K  &  ~~154.535  \pm 1.990\\
f_K/f_{\pi} & ~~ 1.1816  \pm 0.0272\\
f_B &  ~~281.97  \pm 0.39\\
f_{D^+} &  ~~220.1  \pm 0.8\\
f_{D_s^+} &~~ 268.42  \pm 11.22\\
f_+^{K\pi}(0) & ~~ 956.8  \pm 9.1\\
f_+^{B\pi}(0) & ~~ 243.2 \pm 20.8\\
f_+^{D\pi}(0)/f_+^{DK}(0)& ~~  0.833  \pm 0.006\\
{\cal{F}}(1) & ~~ 941.9  \pm 78.7\\
G(1) &  ~~948.1  \pm 149.8\\
 \hline\hline
\end{array}\]\vskip5mm

As one conclusion one can say  that our approach does not confirm  the (approximate) constancy of  $\Delta_R^V $, and at the same time it  allows   a fine structure analysis of all nuclear beta decays, or, more precisely,  of the present procedure for getting a ``constant''  ${\cal{F}}t$. The  solving of the constancy problem  of  $\Delta_R^V $  could require new ideas. 
One suggestion could be the  use of  the present approach, but now with a few more steps.  One can take for a  ``constant'' $\Delta_R^V$ the value given by Eq.\,(\ref{rad1}), as being the best one from all the three, and define a new $\mathcal{F}t_{new}$ as
\begin{eqnarray}
\mathcal{F}t_{new}\equiv ft(1+\Delta_R^V)=\frac{K}{2|G_V|^2(1-\delta_R)(1-\delta_C)}\end{eqnarray}
and try to obtain from fit values for $\delta_R$ and $\delta_C$ , which can be compared with the values computed in \cite{HT1}. A careful analysis  of all these numerical values could say that the values obtained from fit for $\delta_R$ and $\delta_C$ are acceptable, and in such a case the problem is closed, or that they are not compatible with the theoretical knowledge on $T=1$ analog $0^+$ states. In the last case one could think to small contributions due to  scalar and tensor terms in the weak interaction model, or  that $g_V(0)$  has a small nucleus dependence.

We also did a fit by using directly the values for $ft$ given in TABLE IX from \cite{HT}, such that finally we obtained four slightly different matrices for the moduli $V_{ij}$, which were used to get a mean value matrix and its corresponding uncertainty matrix. For that we used the natural embedment of unitary matrices into the double stochastic set, see \cite{PD}, or \cite{PD1}, for details, and the obtained numerical values  for  moduli $V_{ij}$ and their uncertainties are those given in  Table.

Our result on the parameter $g_A/g_V$ was obtained by using practically all the measurements where it is involved. The values spreading is  between 1.25949 corresponding to $A_0$ value from  paper \cite{BY}, and  1.27798 obtained by using neutron lifetime from \cite{WP}, such that the ``unexpected'' Serebrov {\em et al.} result, \cite{AS}, enters naturally in the game.

A second conclusion of this approach is that it allows a ``fine structure analysis'' of all experiments measuring one definite physical quantity, such as $\Delta_R^V$, or $\lambda$, as above, providing to each experimental group one measure of how far from the ideal situation their measured values are standing.

From fit we have obtained also an experimental value on $\delta_c$, which represents  the combined radiative and short-range physics corrections, see \cite{DP}.

The obtained central values for the decay constants and form factors are those normally expected, all the numbers are given in MeV, and the errors are at  $1\sigma$ level. The results provided for  $f_{\pi}$ and  $f_{K}$ are slightly different from that given in \cite{WM}, in particular the lower value for  $f_{K}$.  The most critical situation is that for ${\cal{F}}(1) $ and $G(1)$, whose central values
are almost the same by taking into account the huge errors. We remind that we used the published results, making no scaling as it is usually done, see, e.g., \cite{HFAG}. The simplest situation is that of  $G(1)$ where the older data, \cite{DB} and \cite{MAt}, give a lower value, $G(1)=808.7\pm 77.6$, while the new ones, \cite{JB} and \cite{KAb1}, provide  $G(1)=1,087.4\pm 6.6$, which explain the huge error. The more complicated case is that of ${\cal{F}}(1)$ because of the experimental difficulty to measure the product ${\cal{F}}(1)|U_{cb}|$. For  example, by selecting the highest, and respectively, the lowest values for ${\cal{F}}(1)$,  obtained from data coming  from  \cite{NA}, and respectively \cite{DB}, their ratio is 1.37, which is  too big in our opinion.
 As a matter of fact the highest value  equals $1,133.5$, which is closer to the second value for
$G(1)$ given above. Another source of error are the assumptions made on form factor parameters used to analyse the data, and on the constraints imposed on the shape of these form factors. Of course one can exclude some data, but up to now there is no definite, or accepted by all, procedure to do it.

However these problems can be solved in our approach since the important measurable quantities  from experiment are 
 $|{\cal{F}}^2(\omega_i)\,U_{cb}|^2|$, and these can be used directly in our approach. This procedure can be used to all the  semileptonic decays, and it will lead to  very precise moduli values, and, more important, to  the measurement of form factors  moduli in the physical region. It also allows a direct test of the lepton universality by putting in the fit  the measured data obtained separately for electron, muon and $\tau,$ etc.

More details about our fit, as well as on other physical parameters that can be computed by using the above results, will  be given elsewhere.

As one final conclusion we can say that by taking  properly into account the unitarity constraints we found one tool  which allows the determination of CKM matrix elemnts, and of various decay constants and form factors directly from experimental data.

{\bf Acknowledgements.} We acknowledge a partial support from Program CORINT, CMNP contract no 3/2004.


\begin{thebibliography}{99}

\bibitem{PD} P. Di\c t\u a, J. Math. Phys. {\bf 47},  083510 (2006)

\bibitem{HT} J.C. Hardy and I.S. Towner, Phys.Rev. {\bf C 71}, 055501 (2005)

\bibitem{GS} G Savard {\em et al.}, Phys. Rev. Lett. {\bf 95} 102501 (2005) 
\bibitem{HT1} J. C. Hardy and I. S. Towner, arXiv:0710.3181v1 [nucl-th]

\bibitem{AS} A. Serebrov {\em et al.}, Phys. Lett. B  {\bf 605} 72 (2005)
\bibitem{JN} J. S. Nico {\em et al.}, Phys. Rev. {\bf C 71} 055502 (2005)
\bibitem{SAr} S. Arzumov {\em et al.}, Phys. Lett. B  {\bf 483} 15 (2000)
 \bibitem{JBy} J. Byrne  {\em et al.}, Europhys. Lett. {\bf 33} 187 (1996)
\bibitem{WMa} W. Mampe  {\em et al.}, JETP Lett, {\bf 57} 82 (1993)
\bibitem{VVN} V. V. Nesvizhevskii  {\em et al.}, Sov. Phys.-JETP {\bf 75} 405 (1992)

\bibitem{WMa1} W. Mampe  {\em et al.},  Phys. Rev. Lett. {\bf 63} 593 (1989) 

\bibitem{WP} W. Paul  {\em et al.}, Z. Phys. A {\bf 45} 25 (1989)

\bibitem{PS} P. E. Spivak, Sov. Phys.-JETP {\bf 67} 1735 (1988)
\bibitem{DP} D. Po$\check{\rm c}$ani\'c {\em et al.}, Phys. Rev. Lett. {\bf 93}, 181803 (2004)

\bibitem{WM} W-M. Yao  {\em et al.} [Particle Data Group], J.Phys. {\bf G  33} (2006) 1, and 2007 partial update for edition 2008

\bibitem{IK} K. Ikkado \textit{ et al.} (BELLE Collaboration), Phys. Rev. Lett. {\bf 97}, 251802 (2006)
\bibitem{Ba} B. Aubert   \textit{et al.} (BaBar Collaboration), Phys. Rev.D {\bf 77}, 011107(R)  (2008)
\bibitem{AR} M. Artuso  \textit{at al.} (CLEO Collaboration), Phys. Rev. Lett. {\bf 95} 251801 (2005); hep-ex/0508057 v2

\bibitem{BA} B. Aubert \textit{ et al.} (BaBar Collaboration), Phys. Rev. Lett. {\bf 98} 141801 (2007)
\bibitem{MA} M. Artuso  \textit{et al.} (CLEO Collaboration), Phys. Rev. Lett. {\bf 99} 071802 (2007)
\bibitem{TP} T. K. Pedlar  \textit{et al.} (CLEO Collaboration), Phys. Rev. D {\bf 76}, 072002 (2007)
\bibitem{KAb2} K. Abe  \textit{et al.} (BELLE Collaboration), arXiv:0709.1340v1 [hep-ex]
\bibitem{ME} K. M. Ecklund  \textit{et al.} (CLEO Collaboration), arXiv:
0712.1175v1 [hep-ex]
\bibitem{JR} J. R. Batley  \textit{et al.} (NA48/2 Collaboration), Eur. Phys. J. C {\bf 50}, 329 (2007)
\bibitem{An}A. Antonelli, Nucl.Phys. B(Proc. Suppl) {\bf 167} 18 (2007)
\bibitem{Cb} C. Bloise, Acta Phys.Pol. B {\bf 38} 2731 (2007)
\bibitem{FA} F. Ambrosino \textit{et al.} (KLOE Collaboration), Phys. Lett. {\bf B 632}, 43 (2006)
\bibitem{Mt} M. Testa, Nucl.Phys. B(Proc. Suppl) {\bf 162} 205 (2006)
\bibitem{AA} T. Alexopoulos  \textit{et al.} (KTeV Collaboration), Phys. Rev. Lett. {\bf 93} (2004) 181802

\bibitem{SA} A. Sher \textit{ et al.}, Phys. Rev. Lett. {\bf 91} (2003) 261802
\bibitem{VB} V. Bytev   \textit{et al.}, Eur. Phys. J. C {\bf 27}, 57 (2003)


\bibitem{Vr} V. I. Romanovski \textit{ et al.}, arXiv:0704.2052v1 [hep-ex]
\bibitem{Mm} M. Moulson, arXiv: hep-ex/0703013v1 
\bibitem{Ts} T. Spadaro, arXiv:hep-ex/0703033v1
\bibitem{Ms} M. Antonelli   \textit{et al.} (FlaviaNet Kaon Working Group), arXiv:0801.1817v1 [hep-ex]
\bibitem{FA1} F. Ambrosino  \textit{et al.} (KLOE Collaboration), arXiv: 0802.3009v1 [hep-ex]

\bibitem{JM} J. M. Flynn and J. Nieves, Phys. Rev. D {\bf 76}, 031302(R) (2007)
\bibitem{Bau} B. Aubert  \textit{et al.} (BaBar Collaboration), Phys. Rev. Lett. {\bf 98}, 091801 (2007)
\bibitem{JML} J.M. Link \textit{et al.} (FOCUS Collaboration), Phys. Lett. {\bf B 607}, 233 (2005)

\bibitem{GSH} G.H. Huang \textit{et al.} (CLEO Collaboration), Phys. Rev. Lett. {\bf 94}, 011802 (2005)


\bibitem{DB} D. Buskulic\textit{ et al} (ALEPH Collaboration), Phys. Lett B  {\bf 359} 236 (1995)
\bibitem{DB1} D. Buskulic \textit{ et al} (ALEPH Collaboration), Phys. Lett B  {\bf 395} 373 (1997)
\bibitem{KAc} K. Ackerstaff \textit{ et al} (OPAL Collaboration), Phys. Lett B  {\bf 395} 128 (1997)
\bibitem{GA} G. Abbiendi  \textit{et al} (OPAL Collaboration), Phys. Lett B  {\bf 482} 15 (2000)
\bibitem{PA} P. Abreu  \textit{et al} (DELPHI Collaboration), Phys. Lett B  {\bf 510} 55 (2001)
\bibitem{KAb} K. Abe \textit{et al} (BELLE Collaboration), Phys. Lett. B  {\bf 526} 247 (2002)
\bibitem{NA} N. E. Adam \textit{et al.} (CLEO Collaboration), Phys. Rev. D {\bf 67}, 032002 (2003)
\bibitem{JA} J. Abdallah   \textit{et al.} (DELPHI Collaboration), Eur. Phys. J C {\bf 33} 21 (2004)
\bibitem{AB1} B. Aubert  \textit{et al.} (BaBarCollaboration), Phys. Rev. D {\bf 77}, 032002 (2008)
\bibitem{AB2} B. Aubert \textit{et al.} (BaBarCollaboration),\\  arXiv:0712.3493v2 [hep-ex]

\bibitem{MAt} M. Athanas \textit{et al.} (CLEO Collaboration),  Phys. Rev. Lett. {\bf 79}  2208 (1997)


\bibitem{JB}  J. Bartels\textit{ et al.} (CLEO Collaboration),  Phys. Rev. Lett. {\bf 82} 3746 (1999)
\bibitem{KAb1} K. Abe   \textit{et al.} (BELLE Collaboration), Phys. Lett. B  {\bf 526} 258 (2002)
\bibitem{It} T. M. Ito, arXiv:0704.2365 [nucl-ex]
\bibitem{PB} P. Bopp  \textit{et al.}, Phys. Rev. Lett. {\bf 56} 919 (1986)
\bibitem{Ha} H. Abele \textit{ et al.}, Phys. Lett. B  {\bf 407} 212 (1997)
\bibitem{BY} B. Yerozolimsky   \textit{et al.}, Phys. Lett. B  {\bf 412} 240 (1997)
\bibitem{PL} P. Liaud \textit{ et al.}, Nucl. Phys. A {\bf 612}, 53  (1997)
\bibitem{WJM} W. J. Marciano and A. Sirlin, Phys. Rev. Lett. {\bf 56} 22 (1986)
\bibitem{AS1} A. Sirlin, in \textit{Precision Tests of the Standard Electroweak Model}, edited by P. Langacker (World-Scientific, Singapore, 1994)
\bibitem{PD1} P. Di\c t\u a, Phys. Rev. D {\bf 74}, 113010 (2006)
\bibitem{HFAG} E. Barberio \textit{ et al.}, arXiv:0704.3575v [hep-ex]
\end{thebibliography}
\end{document}